\documentclass[aps,pra,preprint,amsmath,amsfonts,amssymb,a4paper,eqsecnum,superscriptaddress]{revtex4-1}

\usepackage{graphicx}
\usepackage{srcltx}
\usepackage{amsmath}
\usepackage{amsthm}
\usepackage{amsfonts}
\usepackage{amssymb}
\usepackage{amscd}
\usepackage{enumerate}
\usepackage{dsfont}

\usepackage{tikz}
\usetikzlibrary{arrows,backgrounds,snakes}

\tikzstyle{int}=[draw, fill=blue!20, minimum size=2em]
\tikzstyle{init} = [pin edge={to-,thin,black}]

\tikzstyle{abstract}=[rectangle, draw=black, rounded corners, fill=blue!40, drop shadow,
        text centered, anchor=north, text=white, text width=2cm]
\tikzstyle{block} = [rectangle, draw, fill=blue!20, 
    text width=2cm, text centered, rounded corners, minimum height=2em]
\tikzstyle{line} = [draw, -latex']

\def\R{\mathbb{R}}
\def\C{\mathbb{C}}
\def\H{\mathcal{H}}

\def\A{\mathcal{A}}
\def\G{\mathcal{G}}

\def\>{\rightarrow}
\def\-{\mapsto}

\def\p{\psi}

\def\2{\frac{1}{2}}

\def\Tr{\mathrm{Tr}}

\def\be{\begin{equation}}
\def\ee{\end{equation}}
\def\bp{\begin{proof}}
\def\ep{\end{proof}}
\def\bc{\begin{cases}}
\def\ec{\end{cases}}

\newcommand{\bra}[1]{\ensuremath{\left\langle #1\right|}}
\newcommand{\ket}[1]{\ensuremath{\left|#1\right\rangle}}
\newcommand{\braket}[2]{\ensuremath{\left\langle #1\vphantom{#2}\right.\left|\vphantom{#1}#2\right\rangle}}
\newcommand{\tbraket}[2]{\ensuremath{\left\langle #1\vphantom{#2}\right.\left.\otimes\,\,\vphantom{#1}#2\right\rangle}}

\newcommand{\evp}[1]{\ensuremath{\rho(#1)}}

\numberwithin{equation}{section}

\newtheorem{Def}{Definition}[section] 

\newtheorem{The}[Def]{Theorem}

\begin{document}

\markboth{P. Aniello, J. Clemente-Gallardo, G. Marmo and G. F. Volkert}{From Geometric Quantum Mechanics to Quantum Information}

%
%

\title{From Geometric Quantum Mechanics to Quantum Information}

\author{P. Aniello}

\affiliation{Dipartimento di Scienze Fisiche dell'Universit\`a di
  Napoli ``Federico II'', Complesso
  Universitario di Monte Sant'Angelo, via Cintia, I-80126 Napoli,
  Italy}
\email{aniello@na.infn.it//marmo@na.infn.it/volkert@na.infn.it} 
\affiliation{ INFN -- Sezione di Napoli, Complesso Universitario di Monte Sant'Angelo, via Cintia, I-80126 Napoli, Italy}


\author{J. Clemente-Gallardo}

\affiliation{Departamento de F\'{\i}sica Te\'orica, Universidad de Zaragoza,  Campus San Francisco, 50009 Zaragoza (SPAIN)}
\affiliation{BIFI, Universidad de Zaragoza,  Edificio I+D-Campus R\'{\i}o Ebro, Mariano Esquillor s/n, 50018 Zaragoza (SPAIN)}
\email{jesus.clementegallardo@bifi.es}

\author{G. Marmo}
\affiliation{Dipartimento di Scienze Fisiche dell'Universit\`a di
  Napoli ``Federico II'', Complesso
  Universitario di Monte Sant'Angelo, via Cintia, I-80126 Napoli,
  Italy}
\affiliation{ INFN -- Sezione di Napoli, Complesso Universitario di Monte Sant'Angelo, via Cintia, I-80126 Napoli, Italy}
\email{marmo@na.infn.it}

\author{G. F. Volkert}
\affiliation{Dipartimento di Scienze Fisiche dell'Universit\`a di
  Napoli ``Federico II'', Complesso
  Universitario di Monte Sant'Angelo, via Cintia, I-80126 Napoli,
  Italy}



\begin{abstract}
We consider the geometrization of quantum mechanics. We then focus on the pull-back of the Fubini-Study metric tensor field from the projective Hibert space to the orbits of the local unitary groups. An inner product on these tensor fields allows us to obtain functions which are invariant under the considered local unitary groups. This procedure paves the way to an algorithmic approach to the identification of entanglement monotone candidates. Finally, a link between the Fubini-Study metric and a quantum version of the Fisher information metric is discussed.  
\end{abstract}


\maketitle





\section{Introduction}
There are several good reasons for considering a geometrization of quantum mechanics,
as it has been beautifully illustrated in a paper by Ashtekar and Shilling \cite{Ashtekar:1997ud}; consider also the partial list of papers \cite{Heslot:1985,
Rowe:1980,
Cantoni:1975,
Cantoni:1977a,
Cantoni:1977b,
Cantoni:1980,
Cantoni:1985,
Cirelli:1983,
Cirelli:1984,
Abbati:1984,
Provost:1980,
Gibbson:1992,
Brody:2001,
Gosson:2001,
Carinena:2000,
Marmo:2002,
Marmo:2006z,
ClementeGallardo:2007,
Carinena:2007ws,
Grabowski:2000zk,
Grabowski:2005my}, where a geometric formulation of quantum mechanics has been developed. Perhaps, the most appealing reason is provided by the opportunity of making available the whole experience of `classical' methods in the study of quantum mechanical problems. Here we shall focus on some recently established results of the geometrization program of quantum mechanics concerning the study of particular problems of quantum information theory \cite{Aniello:09,Volkert:2010iop,Volkert:2010,Aniello:10, Facchi:2010}.\\
To be more specific, let us comment on what we mean by geometrization of quantum mechanics: To replace the usual Hilbert space picture with a description in terms of Hilbert manifolds, together with all natural implications of this alternative description.\\ In this respect, this proposal is very much similar to the transition from special to general relativity: Space-time is considered to be a Lorentzian manifold and the properties of the Minkowski space time are transferred to the tangent space at each point of the space-time manifold. In particular, we go from the scalar product $\eta_{\mu\nu}X^{\mu}X^{\nu}$ to the Lorentzian metric tensor field $\eta_{\mu\nu}dx^{\mu}\otimes dx^{\nu}$,  which is further generalized to non-flat space-time manifolds in the form   $\eta_{\mu\nu}\theta^{\mu}\otimes\theta^{\nu}$, where $\{\theta^{\mu}\}$ are general 1-forms which carry the information on the non-vanishing of the curvature tensor.\\
Similarly, in the geometrization of quantum mechanics we go from the scalar product $\braket{\p}{\p}$ on the Hilbert space $\H$ to the Hermitian tensor field on the Hilbert manifold, written as $\braket{d\p}{d\p}$.   This would be the associated covariant (0, 2)-tensor field.\\
If we consider as starting carrier space not $\H$ itself but its dual $\H^*$ --- say not ket-vectors but bra-vectors, in Dirac's notation --- we will obtain a (2,0)-tensor field, i.e., a contra-variant tensor field. Once we consider these replacements, algebraic structures will be associated with tensorial structures, and we have to take into account that there will be no more invertible linear transformations but just diffeomorphisms. The linear structure will emerge only at the level of the tangent space and will `reappear' on the manifold carrier space as a choice of each observer \cite{Ercolessi:2007zt}.\\
We must stress that manifold descriptions appear in a natural way already in the standard approach in terms of Hilbert spaces when, due to the probabilistic interpretation of quantum mechanics, we realize that pure states are not vectors in $\H$ but, rather, equivalence classes of vectors, i.e., \emph{rays}. The set of rays, say $\mathcal{R}(\H)$, is the complex projective space associated with $\H$. It is not linear and carries a manifold structure with the tangent space at each point $[\p]$ as `model space'. This space may be identified with the Hilbert subspace of vectors orthogonal to $\psi$.
Other examples of `natural manifolds of quantum states' are provided by the set of density states which do not allow for linear combinations but only convex combinations. They contain submanifolds of density states with fixed rank.\\
The best known example of a manifold of quantum states is provided by the coherent states or any generalized variant \cite{Perelomov:1971,Gilmore:1972,Onofri:1974}, including also non-linear coherent states \cite{Man'ko:1996xv, Aniello:2000}. As is well known, these manifolds of quantum states allow us to describe many properties of the system we are considering by means of finite dimensional smooth manifolds.\\
In this contribution, we start by reviewing, in section \ref{geo of H}, the geometrical formulation of the Hilbert space picture. We shall focus attention on the identification of tensor fields on submanifolds in terms of a natural pull-back procedure as considered in \cite{Aniello:08}. This procedure is applied, in section \ref{app}, by taking account the pull-back on the locally unitarily related quantum states. We then discuss some of its direct consequences for entanglement characterization according to \cite{Aniello:09,Volkert:2010}. In this regard, we relate this tensor fields to the concept of invariant operator valued tensor fields (IOVTs) on Lie groups \cite{Aniello:10}, which naturally admit also applications in the general case of mixed quantum states. In section \ref{QIMetrics}, we review a recently considered connection between the pull-back of the Fubini-Study-metric and a quantum version of the Fisher information metric \cite{Facchi:2010}. We conclude, in section \ref{outlook}, by outlining a relation between IOVTs and the Fisher quantum information metric.


\section{Geometrical Formulation of the Hilbert Space Picture}\label{geo of H}

Consider a separable complex Hilbert space $\H$. A geometrization of this space may be described in two steps as follows. First, by replacing the complex vector space structure with a real manifold $\H^{\R}$, and second, by identifying tensor fields on the latter manifold which are associated with all additional structures being defined on the `initial' Hilbert space, provided by the complex structure, a Hermitian inner product $\braket{\cdot}{\,\cdot}$, Hermitian operators and associated symmetric and anti-symmetric products. Moreover, we'll be interested to focus on geometric structures on  $\H^{\R}$ being defined as pull-back structures from the associated projective Hilbert space of complex rays $\mathcal{R}(\H)$.
\\  
In what follows, our statements should be considered to be always mathematically well defined whenever the Hilbert space we intend to geometrize is finite dimensional. Indeed, the basic ideas coming along the geometric approach in the finite dimensional case are  fundamental for approaching the infinite dimensional case. The additional technicalities which may be required in the latter case will be discussed here by underlining them within specific examples rather than by focusing on general claims (For the manifold point of view for infinite dimensional vector spaces see \cite{Chernoff:1974,Schmid:1987,Lang:1994}).

\subsection{From Hermitian operators to real-valued functions}

Let us start with the identification of tensor fields of order zero. Given a Hermitian operator $A\in u^*(\H)$ defined on a Hilbert space $\H$, we shall find a real symmetric function
\be f_A(\psi) := \braket{\psi}{A\psi},\quad \psi\in \H\ee
on $\H$ and on $\H^{\R}$ respectively. These functions decompose into \emph{elementary quadratic functions}
\be f_{P_j}(\psi) = \braket{\psi}{P_j\psi},\quad \psi\in \H\ee
on $\H^{\R}$ by virtue of a spectral decomposition
\be A= \sum_j\lambda_j P_j\ee
associated with a family of projectors 
$P_j :=\ket{e_j}\bra{e_j}$ and an orthonormal basis $\{\ket{e_j}\}_{j\in I}$ on $\H$. This may be illustrated by taking into account \emph{coordinate functions}
\be  \braket{e_j}{\p}:= z^j(\p) \,.\ee 
yielding 
\be f_{A}(\psi) = \sum_j \lambda_j  f_{P_j}(\psi)=\sum_j \lambda_j|z^j|^2(\p).\ee
In this regard we may recover the eigenvalues and
eigenvectors of the operators at the level of a related function
\be e_A(\psi):=\frac{f_A(\psi)}{\braket{\psi}{ \psi}}\ee
on the punctured Hilbert space $\H_0:=\H-\{0\}$. It is simple to see that eigenvectors are critical points $\psi_*$ of the function $e_A$, i.e.
\be 
de_A(\psi_*)=0 \text{ iff  $\psi_*$ is an eigenvector of $A$.}\ee
Hence,
\be e_A(\psi_*) \text{ is eigenvalue of $A$}.\ee
By virtue of the momentum map
\be \mu: \H_0 \> u^*(\H), \quad \ket{\psi} \mapsto \rho_{\psi}:= \frac{\ket{\psi}\bra{\psi}}{\braket{\psi}{\psi}}\label{m-map}\ee 
we note that \be e_A(\psi)= \rho_{\psi}(A),\quad \rho_{\psi}\in D^1(\H)\ee 
identifies a pull-back function from the set $D^1(\H)$ of normalized rank-1 projectors which are in 1-to-1 correspondence with pure physical states in $\mathcal{R}(\H)$. Hence, $e_A$ is the pull-back of a function which lives on $\mathcal{R}(\H)$.

\subsection{The Fubini-Study metric seen from the Hilbert space}\label{fb-metric}
On this point we shall underline that the momentum map $\mu$, as written within the commutative diagram  
\begin{equation*}
\begin{CD}
\H_0 @>\mu>>u^*(\H)\\
@V\pi VV@A\iota AA\\
\mathcal{R}(\H) @>\cong>>D^1(\H) 
\end{CD}
\end{equation*}
provides a fundamental tool for pulling back, in a \emph{computable} way, any covariant structure defined on $D^1(\H) \cong \mathcal{R}(\H)$
to the `initial' punctured Hilbert space $\H_0$. For this purpose, we may consider for a given Hermitian operator $A$, the operator-valued differential $dA$ in respect to a real parametrization of $u^*(\H)$, and define the $(0,2)$-tensor field
\be \Tr(A dA \otimes dA).\ee 
The differential calculus on a submanifold $\mathcal{M}\subset u^*(\H)$, may then  inherited from the `ambient space' $u^*(\H)$ together with this covariant structure. In particular for $\mathcal{M}\cong \mathcal{R}(\H)$ we find by taking into account the momentum map (\ref{m-map}),   
\be \Tr(\rho_{\psi}d\rho_{\psi}\otimes d\rho_{\psi})
=\frac{\tbraket{d\psi}{d\psi}}{\braket{\psi}{\psi}}- \frac{\braket{\psi}{d\psi}}{\braket{\psi}{\psi}}\otimes \frac{\braket{d\psi}{\psi}}{\braket{\psi}{\psi}}\,,\ee
as momentum-map induced pull-back tensor field on the associated punctured Hilbert space $\H_0$ \cite{Aniello:09}. Moreover, this tensor-field turns out to be identified as a pull-back of the Fubini-Study metric tensor field from the space of rays  $\mathcal{R}(\H)$. Here we shall note that $\ket{d\psi}$ defines a $\H$-vector-valued 1-form which provides a `classical' $\R$-valued 1-form according to $\braket{e_j}{d\psi} \equiv dz_j$, as we shall explain more in detail in the next section. 
 
\subsection{From Hermitian inner products to classical tensor fields}

By introducing an orthonormal basis $\{\ket{e_j}\}_{j\in J}$, we may define coordinate functions on $\H$ by setting
\be \braket{e_j}{\p} = z^j(\psi),\ee 
which we'll write in the following simply as $z^j$. Correspondently, for the dual basis $\{\bra{e_j}\ $ we find coordinate functions \be  \braket{\p}{e_j}= \bar{z}_j(\p^*)\ee
defined on the dual space $\H^*$. By using the inner product we can identify in the finite dimensional case $\H$ and $\H^*$. This provides two possibilities: The scalar product $\braket{\p}{\p}$ gives rise to a covariant Hermitian (0, 2)-metric tensor on $\H$
\be \braket{d\p}{d\p} =\sum_j \braket{d\p}{e_j}\braket{e_j}{d\p}= d\bar{z}_j\otimes dz^j,\ee
where we have used $d\braket{e_j}{\p}=\braket{e_j}{d\p}$, i.e., the chosen basis is not `varied',
or to a contra-variant (2,0) tensor 
\be \braket{\frac{\partial}{\partial \p}}{\frac{\partial}{\partial \p}} = \frac{\partial}{\partial \bar{z}_j}\otimes \frac{\partial}{\partial z^j}\ee
on $\H^*$.
\\\\
\emph{Remark:} Specifically, we assume that an orthonormal basis has been selected once and it does not depend on the base point.
\\\\
By introducing real coordinates, say 
\be z^j({\p}) = x^j(\p)+iy^j(\p)\ee 
one finds       
\begin{eqnarray}
 \braket{d\p}{d\p} = (dx_j \otimes dx^j + dy_j\otimes dy^j)+i(dx_j\otimes dy^j - dy_j\otimes dx^j).  
\end{eqnarray}
Thus the Hermitian tensor decomposes into an Euclidean metric (more generally a Riemannian tensor) and a symplectic form.\\
Similarly, on $\H^*$ we may consider    
\begin{eqnarray}
 \braket{\frac{\partial}{\partial \p}}{\frac{\partial}{\partial \p}}= \bigg(\frac{\partial}{\partial x_j} \otimes \frac{\partial}{\partial x^j} + \frac{\partial}{\partial y_j} \otimes \frac{\partial}{\partial y^j}\bigg)+i\bigg( \frac{\partial}{\partial y_j} \otimes \frac{\partial}{\partial x^j}-\frac{\partial}{\partial x_j} \otimes \frac{\partial}{\partial y^j}\bigg).  
\end{eqnarray}
This tensor field, in contravariant form, may be also considered as a bi-differential operator, i.e., we may define a binary bilinear product on real smooth functions by setting 
\be ((f, g)) = \bigg( \frac{\partial f}{\partial x_j}+i\frac{\partial f}{\partial y_j}\bigg)\cdot  \bigg( \frac{\partial g}{\partial x^j}-i\frac{\partial g}{\partial y^j}\bigg)\ee
which decomposes into a symmetric bracket 
\be (f, g) = \frac{\partial f}{\partial x_j}\frac{\partial g}{\partial x^j} + \frac{\partial f}{\partial y_j}\frac{\partial g}{\partial y^j}\ee
and a skew-symmetric bracket
\be \{f, g\} = \frac{\partial f}{\partial y_j}\frac{\partial g}{\partial x^j} - \frac{\partial f}{\partial x_j}\frac{\partial g}{\partial y^j}.\ee
This last bracket defines a Poisson bracket on smooth functions defined on $\H$.\\
Summarizing, we can replace our original Hilbert space with an Hilbert manifold, i.e. an even dimensional real manifold on which we have tensor fields in covariant form 
\be G = dx_j\otimes dx^j + dy_j\otimes dy^j\ee 
\be \Omega = dy_j\otimes dx^j -  dx_j\otimes dy^j,\ee
or tensor fields in contravariant form
\be G^{-1} = \frac{\partial}{\partial x_j} \otimes \frac{\partial}{\partial x^j} + \frac{\partial}{\partial y_j} \otimes \frac{\partial}{\partial y^j}\ee
\be \Omega^{-1} = \frac{\partial}{\partial y_j} \otimes \frac{\partial}{\partial x^j}-\frac{\partial}{\partial x_j} \otimes \frac{\partial}{\partial y^j},\ee 
along with a complex structure tensor field
\be J = dx^j\otimes \frac{\partial}{\partial y_j}-dy^j\otimes \frac{\partial}{\partial x^j}.\ee 
The contravariant tensor fields, considered as bi-differential operators define a symmetric product and a skew symmetric product on real smooth functions. The skew-symmetric product actually defines a Poisson bracket. In particular, for functions 
\be f_{A}(\psi) = \braket{\psi}{A\psi},\quad \psi\in \H,\ee
associated with Hermitian operators $A$, we shall end up with the relations
\be f_{[A, B]_+} \equiv G^{-1}(df_A, df_B).\ee
\be f_{[A, B]_-} \equiv \Omega^{-1}(df_A, df_B),\ee
which replaces symmetric and anti-symmetric operator products $[A, B]_{\pm}$ by symmetric and anti-symmetric tensor fields respectively. Hence, via these tensor fields we may identify symmetric and Poisson brackets on the set of quadratic functions according to 
\be f_{[A, B]_+} = (f_A, f_B),\ee
\be f_{[A, B]_-} = \{f_A, f_B\},\ee
which synthesize to a star-product
\be  ((f, g)) = (f_A, f_B)+i\{f_A, f_B\}:=f_A\star f_B \ee
and turn therefore the set of quadratic functions into a C-star algebra. In this way we may encode the original non-commutative structure on operators in terms of `classical', i.e.\,Riemannian and symplectic tensor fields according to
\be f_A\star f_B=f_{A\cdot B}(\psi) = (G^{-1}+i\Omega^{-1})(df_A(\psi), df_B(\psi)).\ee
To take into account the geometry of the set of physical (pure) states,  we need to modify $G^{-1}$ and
$\Omega^{-1}$ by a conformal factor to turn them into projectable tensor fields on $\mathcal{R}(\H)$. The projection is generated at the infinitesimal level by the real and imaginary parts of the action
of $\C_0$ on $\H_0$ given by the dilation vector field $\Delta$ and the
$U(1)$-phase rotation generating vector field $\Gamma:=J(\Delta)$ respectively.  In this way we shall identify 
\be \widetilde{G}(\p)=\braket{\p}{\p}G^{-1}-(\Delta\otimes \Delta+\Gamma\otimes \Gamma)\ee
\be \widetilde{\Omega}(\p)=\braket{\p}{\p}\Omega^{-1}-(\Delta\otimes \Gamma-\Gamma\otimes \Delta),\ee
as projectable structures \cite{Chruscinski:2008}. They establish a Lie-Jordan algebra structure on the space of
real valued functions whose Hamiltonian vector fields are also Killing vector
fields for the projection $\tilde G$. In this regard one finds
a generic function on $\mathcal{R}(\H)$ defines a quantum evolution, via the associated Hamiltonian vector field, if and only if the vector field is a derivation for the Riemann-Jordan product \cite{Cirelli:1991, Marmo:2006z}.
\\
The geometric formulation of the Hilbert space picture reviewed here so far can be summerized at this point by a `dictionary' as follows \cite{Heslot:1985,
Rowe:1980,
Cantoni:1975,
Cantoni:1977a,
Cantoni:1977b,
Cantoni:1980, 
Cantoni:1985,
Cirelli:1983, 
Cirelli:1984,
Abbati:1984,
Provost:1980,
Ashtekar:1997ud,
Gibbson:1992,
Brody:2001,
Gosson:2001,
Carinena:2000,  
Marmo:2002,
Marmo:2006z,
ClementeGallardo:2007,
Carinena:2007ws,
Grabowski:2000zk,
Grabowski:2005my}.
\begin{center}
\begin{tabular}{|c|l|}
\hline
\textsc{Standard QM} & \textsc{Geometric QM}\\
\hline
\hline
Complex vector space  & Real manifold with a complex structure\\
Hermitian inner product $\langle\cdot ,\cdot \rangle$ & Hermitian tensor field\\
Real part of $\langle\cdot ,\cdot \rangle$ & Riemannian tensor field \\
Imaginary part of $\langle\cdot ,\cdot \rangle$ & Symplectic tensor field \\
Hermitian operator $A$  & Real-valued function $e_A(\psi):=\frac{\langle\psi , A\psi \rangle}{\langle\psi ,\psi\rangle}$ \\
Eigenvectors of $A$ & Critical points of $e_A(\psi)$\\
Eigenvalues of $A$ & Values of $e_A$ at critical points\\
Commutator & Poisson bracket\\ 
Anti-commutator & Symmetric bracket\\ 
Quantum evolution & Hamiltonian Killing vector field\\
\hline
\end{tabular} 
\end{center} 

\subsection{Pull-back structures on submanifolds of $\H$}\label{pb from H}
One interesting aspect for the current applications of the geometric formulation of quantum mechanics is the possibility to induce tensor fields in covariant form on a given submanifold via a pull-back procedure \cite{Aniello:08, Aniello:09,Volkert:2010}. In particular one finds 
\begin{The}\label{BP-Th} Let $\{\theta_j\}_{j\in J}$ be a basis of left-invariant 1-forms on a Lie group $\mathcal{G}$, and let
$\{X_j\}_{j\in J}$ be a dual basis of left-invariant vector fields, and let $iR$ be the infinitesimal representation of $U:\G\rightarrow U(\H)$, inducing for $\ket{\psi}\in S(\H)$ a map
$$f_{\G}:\G\rightarrow \H,$$ $$f_{\G}(g):=U(g)\ket{\psi},$$
and let $$\sum_{j=1}^N d\bar{z}^{j}\otimes dz^{j}=\sum_{j=1}^N\underbrace{dx^{j}\odot dy^{j}+dx^{j}\odot dy^{j}}_{:=G}+i\underbrace{(dx^{j}\wedge dy^{j})}_{:=\Omega }$$
be an invariant Hermitian tensor field on $\H\cong \C^N\cong \R^{2N}$. Then
$$f_{\G}^*(\sum_{j=1}^N d\bar{z}^{j}\otimes dz^{j})=\rho^{\psi}(R(X_j)R(X_k)) \theta^{j} \otimes \theta^{k}:=T_{\G}^{\rho^{\psi}} $$
for $\rho^{\psi} := \frac{\ket{\psi}\bra{\psi}}{\braket{\psi}{\psi}} \in D^1(\H)$. 
\end{The}
As a direct consequence, we shall identify this (degenerate) pull-back tensor field with the pull-back of a non-degenerate pull-back tensor field which lives on a homogenous space $\G/\G_0$. The latter admits a smooth  embedding via the unitary action of the Lie Group as orbit manifold $\mathcal{O}$ in the Hilbert space and establishes therefore a pull-back of the Hermitian structure both on the orbit $\mathcal{O}$ and the homogenous space $\G/\G_0$. Hence, the computation of the pull-back on the orbit, reduces to the the computation of the pull-back on the Lie group, as indicated here in the commutative diagram below.
\begin{equation*}
\begin{CD}
\G @>f_{\G}>>\H\\
@V\pi VV@A\iota AA\\
\G/\G_0 @>\cong>>\mathcal{O}, 
\end{CD}
\end{equation*}
where $\pi$ denotes the canonical projection of $\G$ onto $\G/\G_0$ and $\iota$ defines the inclusion map of the orbit $\mathcal{O}$ on $\H$.\\
Taking into account in this regard the space of pure states provided by the projective Hilbert space $\mathcal{R}(\H)$, it becomes appropriated to consider
the covariant tensor field 
\be \frac{d\bar{z}^{j}\otimes dz^{j}}{\sum_j |z^{j}|^2}-\frac{z^{j}d\bar{z}^{j}\otimes \bar{z}^{k}dz^{k}}{(\sum_j |z^{j}|^2)^2}\label{ProjectiveHT}\ee
on $\H_0$ which has been identified in section \ref{fb-metric} as pull-back tensor of the Fubini study metric from $\mathcal{R}(\H)\cong \C P^{n}$ to $\H_0\cong \C^{n+1}_0$. Here the pull-back on the Lie group reads
\be (\rho^{\psi}(R(X_j)R(X_k))-  \rho^{\psi}(R(X_j))\rho^{\psi}(R(X_k))\theta^j\otimes \theta^k.\label{ProjectivePBTonG}\ee 
The embedding of the Lie group and its corresponding orbit is related to the co-adjoint action map on all group elements modulo U(1)-representations $U(h)=e^{i\phi(h)}$
\be f_{\G}^{U(1)}: \G/U(1) \rightarrow \mathcal{R}(\H), \quad g\mapsto U(g)\rho U(g)^{\dagger},\quad \rho\in \mathcal{R}(\H).\ee
Let us underline again that the structure (\ref{ProjectivePBTonG}) is defined \emph{on the Lie group via a pull-back tensor field from the Hilbert space} even though it contains the full information of the (non-degenerate) tensor field on the corresponding co-adjoint orbit $\mathcal{O}$ which is embedded in the projective Hilbert space. The additional $U(1)$- degeneracy is here captured in a corresponding enlarged isotropy group $\G_0^{U(1)}$ according the commutative diagram below.
\begin{equation*}
\begin{CD}
\G @>f_{\G}>>S(\H)\\
@VU(1) VV@VU(1) VV\\
\G/U(1) @>f_{\G}^{U(1)}>>\mathcal{R}(\H)\\ 
@V\pi VV@A\iota AA\\
\G/\G_0^{U(1)} @>\cong>>\mathcal{O}
\end{CD}
\end{equation*} 
This approach provides therefore in an `algorithmic' procedure to find a geometric description of coherent state manifolds, as defined in \cite{Perelomov:1971,Gilmore:1972,Onofri:1974}. Indeed, the associated orbits in our approach turn out to be more general as those give by coherent states, whenever we allow to take into account also reducible representations, as it typically occurs in composite Hilbert spaces. 

\section{Some Applications :
Composite Systems, Entanglement and Separability}\label{app}

\subsection{Separable and maximal entangled pure states}
By considering the representation
\begin{align}
\mathcal{G}\equiv U(n)\times U(n) \rightarrow &  U(n^2)\notag\\
 g\equiv (g_A, g_B) \mapsto & U(g)\equiv g_A\otimes g_B =( g_A\otimes \mathds{1})(\mathds{1}\otimes  g_{B})\end{align}
infinitesimal generated by generalized Pauli-matrices tensored by the identity of a subsystem 
\begin{align}
\mbox{Lie}(\mathcal{G}) \equiv u(n)\oplus u(n) \rightarrow &  u(n^2)\notag\\
 X_j \mapsto & iR(X_j) \equiv 
 \begin{cases}
 i\sigma_j\otimes \mathds{1} &\text{ for } 1 \le j \le n^2\\
 \mathds{1}\otimes  i\sigma_{j-n^2} &\text{ for } n^2+1 \le j \le 2n^2,
\end{cases}\end{align} 
one finds according to theorem \ref{BP-Th} a pull-back tensor field on the Lie group  
$$f_{\G}^*(\delta_{jk}d\bar{z}^{j}\otimes dz^{k}) = \rho^{\psi}(R(X_j)R(X_k)) \theta^{j} \otimes \theta^{k}$$ 
$$=\underbrace{\rho^{\psi}([R(X_j)R(X_k)]_+) \theta^{j} \odot \theta^{k}}_{=f_{\G}^*G}+i\underbrace{\rho^{\psi}([R(X_j)R(X_k)]_-) \theta^{j} \wedge \theta^{k}}_{=f_{\G}^*\Omega}$$ 
which decomposes for all $\rho^{\psi} \in D^1(\C^n\otimes \C^n)$ into a Riemannian and a symplectic coefficient matrix 
\be (T^{\rho^{\psi}}_{jk})= \left(\begin{array}{cc}(A^{\rho_A}_{(jk)}) & (C^{\rho^{\psi}}_{jk})  \\(C^{\rho^{\psi}}_{jk})  & (B^{\rho_B}_{(jk)}) \end{array}\right)+i  \left(\begin{array}{cc}(A^{\rho_A}_{[jk]}) & 0 \\0  & (B^{\rho_B}_{[jk]}) \end{array}\right),\notag\ee
\begin{align}
A^{\rho_A}_{(jk)}= &  \evp{[\sigma_j,\sigma_k]_+\otimes \mathds{1}}=\rho_A([\sigma_j,\sigma_k]_+)\\\notag\\
A^{\rho_A}_{[jk]}= &  \evp{[\sigma_j,\sigma_k]_-\otimes \mathds{1}} =\rho_A([\sigma_j,\sigma_k]_-)\\\notag\\
C^{\rho^{\psi}}_{jk}= & \evp{\sigma_j\otimes \sigma_{k-n^2}}.\end{align}
In contrast to the Riemannian part, we observe that the symplectic part splits in general into two symplectic structures associated with the subsystems. Hence, the symplectic structure behaves in analogy to classical composite systems. This may suggest to consider following definition and associated theorem \cite{Aniello:10}: 
\begin{Def} $\rho^{\psi} \in D^1(\C^n\otimes \C^n)$ is called \textit{maximally entangled}
if $$f_{U(n)\times U(n)}^*\Omega=0.$$
\end{Def}
Based on this definition, we find
\begin{The} $\rho^{\psi} \in D^1(\C^n\otimes \C^n)$ is a maximally entangled iff the reduced state is maximally mixed.
\end{The}
Hence, this theorem recovers the definition \cite{Donald:2002} which provides the von Neumann entropy as the unique measure of entanglement for pure bi-partite states.\\
On the other extreme, we find for separable states a factorization of the Riemannian coefficient sub-matrix $C$ into reduced density states according to  
$$ \rho^{\psi} \mbox{ is \textit{separable} }\Leftrightarrow C^{\rho^{\psi}}_{jk} = \rho^{\psi}(\sigma_j\otimes \sigma_{k-n^2}) = \rho_A(\sigma_j)\rho_B(\sigma_{k-n^2}).$$
In contrast, if we take the pull-back tensor field
\be \rho^{\psi}(R(X_j)R(X_k)) \theta^{j} \otimes \theta^{k}- \rho^{\psi}(R(X_j))\rho^{\psi}(R(X_k)) \theta^{j} \otimes \theta^{k}:=\mathcal{T}^{\rho^{\psi}}_{\G} \ee
provided by the Fubini-Study metric from the projective Hilbert space we find the modified coefficient sub-matrix
\be  \mathcal{C}^{\rho^{\psi}}_{jk} := \rho^{\psi}(\sigma_j\otimes \sigma_{k-n^2}) - \rho_A(\sigma_j)\rho_B(\sigma_{k-n^2}),\ee
and therefore a splitting condition
$$ \rho^{\psi} \mbox{ is \textit{separable} }
\Leftrightarrow \mathcal{T}^{\rho^{\psi}}_{U(n)\times U(n)}=\mathcal{T}^{\rho_A\otimes \rho_B}_{U(n)\times U(n)} = \mathcal{T}^{\rho_A}_{U(n)} \oplus \mathcal{T}^{\rho_B}_{U(n)}.$$ 
Hence, we may detect separable states, as those provided by a Segre-embedding
\be \mathcal{R}(\H_A)\times \mathcal{R}(\H_B) \hookrightarrow \mathcal{R}(\H_A\otimes \H_B)\ee
seen from the Hilbert space by the condition $\mathcal{C}^{\rho^{\psi}}_{jk}=0$.

\subsection{Quantitative statements}

For approaching in this setting quantitative statements we may consider invariant functions
$$f(\p) := \braket{\mathcal{T}^{\rho^{\psi}}_{U(n)\times U(n)}}{\mathcal{T}^{\rho^{\psi}}_{U(n)\times U(n)}}$$ 
under local unitary transformations provided by a Hermitian inner product on invariant tensor fields on $U(n)\times U(n)$ associated with the pullback of the Fubini-Study metric seen from the Hilbert space. More specific, we find
$$\mathcal{T}^{\rho^{\psi}}_{\G}=\mathcal{T}^{\rho^{\psi}}_{j_1j_2}\theta^{j_1}\otimes \theta^{j_2}$$
$$\braket{\mathcal{T}^{\rho^{\psi}}_{\G}}{\mathcal{T}^{\rho^{\psi}}_{\G}}:=(\mathcal{T}^{\rho^{\psi}}_{j_1j_2})^*\mathcal{T}^{\rho^{\psi}}_{k_1k_2}\braket{\theta^{j_1}\otimes \theta^{j_2}}{\theta^{k_1}\otimes \theta^{k_2}}.$$
With  $\braket{\theta^j}{\theta^k}= \delta^{jk}$  this gives rise to
$$\braket{\mathcal{T}^{\rho^{\psi}}_{\G}}{\mathcal{T}^{\rho^{\psi}}_{\G}}=\sum_{j_1,j_2}|\mathcal{T}^{\rho^{\psi}}_{j_1j_2}|^2:=\|\mathcal{T}^{\rho^{\psi}}_{_{j_1j_2}} \|^2_2.$$
In particular, we may consider an inner product on the symmetric part
$$\|\mathcal{T}^{\rho^{\psi}}_{_{(jk)}} \|^2_2=\|\mathcal{A}^{\rho_A}_{_{(jk)}} \|^2_2+\|\mathcal{B}^{\rho_B}_{_{(jk)}} \|^2_2+2\|\mathcal{C}^{\rho^{\psi}} \|^2_2$$
which implies an entanglement monotone candidate,\emph{ which evades the explicit computation of Schmidt-coefficients} (compare also \cite{Schlienz:1995,Man'ko:2002ti}). In particular, we find \cite{Aniello:09,Volkert:2010}
\begin{The} Let $\rho^{\psi}\in D^1(\C^n\otimes \C^n)$ and let $\rho_A, \rho_B\in D(\C^n)$ be the reduced density states of $\rho$. Then 
$$\frac{1}{n^2}\|\mathcal{C}^{\rho^{\psi}} \|_2 = \|\rho-\rho_A\otimes \rho_B\|_2.$$
\end{The}

\subsection{Mixed states entanglement and invariant operator valued tensor fields}

\begin{figure}[htp]
\centerline{\includegraphics[scale=0.50]{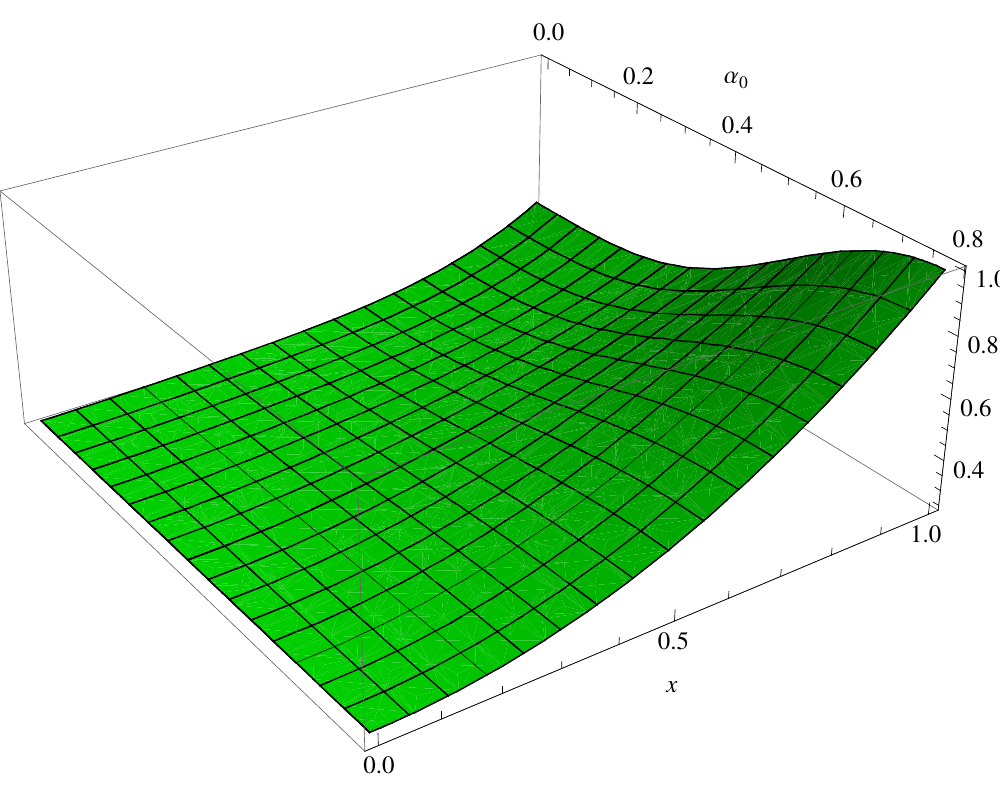}\,\,\includegraphics[scale=0.50]{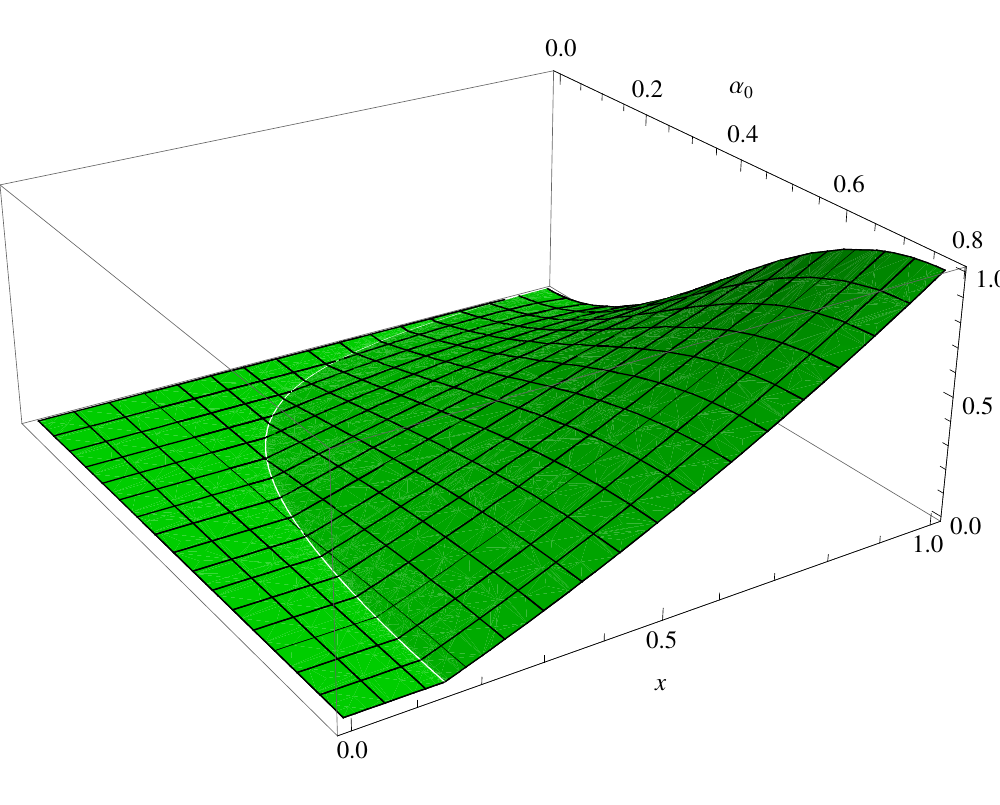}}
\vspace*{8pt}
\caption{The function $f_2^{\tilde{R}}(\rho)$ gives rise to a possible approximation (left) to the concurrence measure (right) applied to a family of 2-parameter states on a composite Hilbert space of two qubits. \label{fig1}}
\end{figure}

So far we modeled an entanglement characterization \emph{algorithm} based on invariant tensor fields on the Lie group $\G=U(n)\times U(n)$, which `replaces' functions on Schmidt-coefficients by functions on tensor-coefficients:\begin{center}
\begin{tikzpicture}[node distance=3.5cm,auto,>=latex']
    \node [int, pin={[init]above:$\rho^{\psi}$}] (a) {$\mathcal{T}^{\rho^{\psi}}_{\G}$};
    \node (b) [left of=a,node distance=2cm, coordinate] {$ $};
    \node  (c) [right of=a] {$f(\p) :=\sum_{j_1j_2}|\mathcal{T}^{\rho^{\psi}}_{j_1j_2}|^2$};
    \node [coordinate] (end) [right of=c, node distance=2cm]{};
    \path[->] (b) edge node {$(R(X_j))$} (a);
    \path[->] (a) edge node {$ $} (c);  
\end{tikzpicture}\end{center}
Similar to the case of pure states, we shall also identify in the generalized regime of mixed states entanglement monotone candidates by functions 
\be f:D(\H_A\otimes \H_B) \rightarrow \R_+ \ee 
which are invariant under the local unitary group of transformations $U(\H_A)\times U(\H_B)$ \cite{Vidal:2000}. In this necessary strength, we propose in the following entanglement monotones candidates by taking into account constant functions on local unitary orbits of entangled quantum states, arising from invariant operator valued tensor fields (IOVTs) on $U(\H_A)\times U(\H_B)$ as considered recently on general matrix Lie groups $\G$ \cite{Aniello:10}. Let us review the basic construction.\\
Given a unitary representations
\be U: \mathcal{G}\rightarrow U(\H),\ee
we may identify an anti-Hermitian operator-valued left-invariant 1-form
\be-U(g)^{-1}dU(g)\equiv iR(X_j)\theta^j\ee
on $\G$, where the operator $iR(X_j)$ is associated with the representation of the Lie algebra $\text{Lie}(\mathcal{G})$.  In this way, we may construct higher order invariant operator valued tensor fields 
\be -U(g)^{-1}dU(g) \otimes U(g)^{-1}dU(g)=R(X_j)R(X_k)\theta^j\otimes \theta^j,\ee
on $\mathcal{G}$ by taking into account the representation as being equivalently defined by means of the representation of the enveloping algebra of the Lie algebra in the operator algebra $\A :=$End$(\H)$. More specific, any element $X_j\otimes X_k$ in the enveloping algebra becomes associated with a product 
\be R(X_j)R(X_k)\in \A:=\mbox{End}(\H),\ee
where $\A$, may denote the vector space of a $C^*$-algebra. On this point, we may evaluate each one of these products by means of dual elements \be \rho\in \A^*,\ee
according to
\be \rho(R(X_j)R(X_k)) \equiv \Tr(\rho\, R(X_j)R(X_k))\in \C,\ee
yielding a complex-valued tensor field
\be \rho(R(X_j)R(X_k))\theta^j\otimes \theta^j\label{rho tensor}\ee
on the group manifold. By taking the k-th product of invariant operator-valued left-invariant 1-forms
\be -U(g)^{-1}dU(g) \otimes U(g)^{-1}dU(g)\otimes...\otimes U(g)^{-1}dU(g),\ee
we shall find a representation R-dependent IVOT of order $k$
\be \theta_R := \bigg(\prod_{a=1}^k R(X_{i_a})\bigg) \bigotimes_{a=1}^k\theta^{i_a}\notag\ee 
on a Lie group $\G=U(n)\times U(n)$. After evaluating it with a \emph{mixed} quantum state 
\be \theta_R \mapsto \rho(\theta_R):=\theta^{\rho}_R =\rho\bigg(\prod_{a=1}^k R(X_{i_a})\bigg) \bigotimes_{a=1}^k\theta^{i_a}\notag\ee
one may again consider invariant functions via an inner product $\braket{\theta^{\rho}_R}{\theta^{\rho}_R}$. In particular, for $k=n=2$, we recover in this way the purity and the concurrence related measures involving a spin-flip transformed state $\tilde{\rho}$ by considering inner product combinations of symmetric and anti-symmetric tensor fields 
\be G^{\rho}_R :=\rho([R(X_j),R(X_k)]_+),\quad  \Omega^{\rho}_R :=\rho([R(X_j),R(X_k)]_-),\ee
according to 
\be \frac{1}{8}\big(\braket{G^{\rho}_R}{G^{\rho}_R}+ (-1)^s\braket{\Omega^{\rho}_R}{\Omega^{\rho}_R}\big) -\frac{1}{2}= \begin{cases}
 \mbox{Tr}(\rho^2) &\text{ for } s=0\\
 \mbox{Tr}(\rho\tilde{\rho}) &\text{ for } s=1. 
\end{cases}.\ee  
In more general terms, one may introduce $R$-\emph{classes} of entanglement monotone candidates by taking into account polynomials 
$$f_k^{R}(\rho):=\sum_n a_n\braket{\theta^{\rho}_R}{\theta^{\rho}_R}^n, \quad \theta^{\rho}_R :=\rho\bigg(\prod_{a=1}^k R(X_{i_a})\bigg) \bigotimes_{a=1}^k\theta^{i_a}. $$
The case
\be \tilde{R}(X_j)= R(X_j)- \rho(R(X_j))\mathds{1},\label{NL-op}\ee
recovers for IOVTs of order $k=2$, a class of separability criteria associated with covariance matrices (CMs) (Gittsovich et al. 2008) by means of a \emph{CM-tensor field} \be\theta^{\rho}_{\tilde{R}}=(\rho(R(X_j)R(X_k))-  \rho(R(X_j))\rho(R(X_k))\theta^j\otimes \theta^k.\ee
An open problem in the field of CM-ctiteria is provided by the question how to find an extension to quantitative statements \cite{Gittsovich:2008}. A possible approach could be provided here by taking into account a $\tilde{R}$-\emph{class} of entanglement monotone-candidates by considering $$f_2^{\tilde{R}}(\rho)= \sum_n a_n\braket{\theta^{\rho}_{\tilde{R}}}{\theta^{\rho}_{\tilde{R}}}^n.$$
To give an example, we consider the function
\be f_2^{\tilde{R}}(\rho)\equiv \braket{\theta^{\rho}_{\tilde{R}}}{\theta^{\rho}_{\tilde{R}}}\ee
applied to a family of 2-parameter states on a composite Hilbert space of two qubits given by   
\be \rho_{x, \alpha_0}:= x\ket{\alpha_0} \bra{\alpha_0} +(1-x)\rho^*, \quad \ket{\alpha_0}:=\cos(\alpha_0)\ket{11}+\sin(\alpha_0)\ket{00}\ee 
and find a possible approximation to the concurrence measure\\
\be\mbox{max}[\lambda_4-\lambda_3-\lambda_2-\lambda_1, 0], \quad\lambda_j\in\mbox{Spec}(\rho\widetilde{\rho}).\ee
Both functions are plotted in figure \ref{fig1}.

\section{From quantum to classical information}\label{QIMetrics}

In the previous section we considered invariant operator valued tensor fields (IOVTs) on Lie groups to tackle the problem of entanglement quantification in composite quantum systems. As a source for performing quantum computation, quantum communication and other types of quantum information processes, we may ask how the resulting entanglement monotone candidates which we have discussed so far are related to known quantum information measure, in analogy to the von Neumann entropy
$$S(\rho)=-\Tr(\rho \log \rho),$$
which establishes a unique entanglement measure for pure states, when applied to corresponding reduced density states \cite{Donald:2002}. In this regard we may consider the \emph{quantum relative entropy}
$$S(\rho||\rho')=S(\rho)-\Tr(\rho \log \rho'),$$
which introduces the notion of a distance between quantum states. In particular, it defines a distance which is monotone under completely positive maps $\Phi$ \cite{Bengtsson:2006},  
$$S(\Phi \rho||\Phi \rho')\leqslant S(\rho||\rho').$$
More general \cite{Gibilisco:2007}, a completely positive map-monotone metric on the space of quantum states $D(\H)$ may establish the notion of a \emph{quantum Fisher information metric}. It is of general interest to understand under which condition the \emph{classical} Fisher information metric can be recovered from a given quantum information metric. In contrast to the latter, we shall note that the classical Fisher metric is uniquely defined as a Markov map-monotone metric on the space of classical probability distributions.\\ 
A frequently used quantum Fisher information metric on a real submanifold of quantum states 
\be \rho_{\theta}\in \mathcal{N}\subset D(\H),\label{theta-sm rho}\ee
parametrized by $\theta\in \R^{\dim(\mathcal{N})}$, is given by 
\be I(\theta):= \Tr(\rho_{\theta}d_{l}\rho_{\theta}\otimes d_{l}\rho_{\theta})\label{QIFM}\ee
with the implicitly defined logarithmic differential $d_{l}$ related to an operator-valued 1-form
\be d\rho_{\theta}=\frac{1}{2}(\rho_{\theta} d_{l}\rho_{\theta}+d_{l}\rho_{\theta}\rho_{\theta}),\ee
where the `ordinary' differential $d\rho_{\theta}$ is considered in respect to the parameters $\theta$.\\
In the case of pure states $\rho=\rho^2$, one finds
\be d\rho^2_{\theta}=\rho_{\theta} d\rho_{\theta} +d\rho_{\theta} \rho_{\theta} =d\rho_{\theta},\ee
and therefore $d_l \rho_{\theta} = 2d\rho_{\theta}$. In conclusion,
\be I(\theta)= 4\Tr(\rho_{\theta}d\rho_{\theta}\otimes d\rho_{\theta})\quad \mbox{if $\rho_{\theta}$ is pure.}\ee
By taking into account the pull-back induced by the momentum map on the associated Hilbert space according to section \ref{fb-metric}, 
one may identify a submanifold $\mathcal{M}\subset \H_0$ of Hilbert space vectors  $\ket{\psi_{\theta}}\in \mathcal{M}$, such that the restriction of the momentum map
\be \mu|_{\mathcal{M}}:\mathcal{M} \rightarrow u^*(\H),\quad \ket{\psi_{\theta}} \mapsto\mu(\ket{\psi_{\theta}})=\frac{\ket{\psi_{\theta}}\bra{\psi_{\theta}}}{\braket{\psi_{\theta}}{\psi_{\theta}}}:=\rho^{\psi}_{\theta}\ee
on this submanifold gives rise to a pullback tensor field 
$$ K:=\Tr(\rho^{\psi}_{\theta}d\rho^{\psi}_{\theta}\otimes d\rho^{\psi}_{\theta})$$
\be=\frac{\tbraket{d\psi_{\theta}}{d\psi_{\theta}}}{\braket{\psi_{\theta}}{\psi_{\theta}}}- \frac{\braket{\psi_{\theta}}{d\psi_{\theta}}}{\braket{\psi_{\theta}}{\psi_{\theta}}}\otimes \frac{\braket{d\psi_{\theta}}{\psi_{\theta}}}{\braket{\psi_{\theta}}{\psi_{\theta}}}\,,\label{PHT}\ee
on $\mathcal{M}\subset \H_0$. Hence, we have the pull-back of the Fubiny Study metric tensor field from the space of rays  $\mathcal{R}(\H)$ to $\mathcal{N}$, seen from a submanifold $\mathcal{M}$ in the Hilbert space. As a consequence, and in accordance to \cite{Facchi:2010}, the pull-back tensor field $K$ on a submanifold $\mathcal{M}$ of quantum state vectors 
\be \psi(x,\theta) \equiv p(x, \theta)^{1/2} e^{W(x, \theta)}\in L^2(\R^n),\quad x\in \R^n,\quad  \theta\in \R^{\dim(\mathcal{N})}\label{theta-sm},\ee
is related to the quantum information metric $I(\theta)$ in 
(\ref{QIFM}) if $\rho_{\theta}$ is a pure state.\\
To illustrate the pull-back, we define for any given tensor field $T(x,\theta)$ of order $r$ (including functions for order $r=0$) the generalized expectation value integral
\be \mathbb{E}_p(T):= \int_{\R^n} p(x,\theta) T(x,\theta) dx,\label{deriv1}  \ee
which `traces out' the $x$-dependence of the tensor field $T$. A straightforward computation (see \cite{Facchi:2010}) yields then the identification of a pull-back tensor field 
\be K= G +i\Omega,\ee  
on the submanifold $\mathcal{M}$ which is decomposed into a symmetric tensor field 
\be G:= \mathbb{E}_p((d\ln p)^{\otimes 2}) +\mathbb{E}_p(dW ^{\otimes 2})- \mathbb{E}_p(dW)^2\label{f1}\ee 
and a antisymmetric tensor field 
\be \Omega := \mathbb{E}_p(d\ln p\wedge dW).\ee
Moreover, by taking into account in the symmetric part a further decomposition
\be G\equiv \mathcal{F} + \mbox{Cov}(dW),\ee
one recovers the \emph{Classical Fisher Information metric}
\be \mathcal{F} := \mathbb{E}_p((d\ln p)^{\otimes 2})\ee
and a \emph{phase-covariance matrix tensor field} 
 \be \mbox{Cov}(dW):=\mathbb{E}_p(dW ^{\otimes 2})- \mathbb{E}_p(dW)^2.\label{f2}\ee
For the parts of the pull-back tensor field containing the phase $W$ in differential form, we may therefore identify for pure states according to the non-classical counterpart of the Fisher classical within the quantum information metric. As a matter of fact,
the quantum information metric collapses to the \emph{classical Fisher information metric} for \be dW=0,\ee
i.e.\,if the phase is constant.

\section{Conclusions and Outlook}\label{outlook}

For a given embedding of the manifold $\mathcal{M}$ into the Hilbert space $\H$,
\be f:\mathcal{M}\rightarrow \H,\ee
we have seen that, for pure states $\rho\in D^1(\H)$,
\be I(\theta) = f^*_{\mathcal{M}}\Tr(\rho d\rho\otimes \rho),\ee
while, for IOVTs associated with the realization (\ref{NL-op}), evaluated on pure states we have
\be \rho(\mbox{IOVT}) = f^*_{\G}\Tr(\rho d\rho\otimes \rho).\ee
But then we have
\be I(\theta) =  \rho(\mbox{IOVT})  \mbox{ if } \mathcal{M}=\G.\ee
Hence, pure quantum state-evaluated IOVTs are directly related to a quantum information metric, if the pull-back of $\Tr(\rho d\rho\otimes \rho)$ is made on a Lie group and  associated $\G$-orbits respectively.\\
To identify pull-back tensor fields from $\mathcal{R}(\H)$ associated with mixed quantum states,
we shall consider the pull-back on $$\G\equiv U(n)\times U(n)\times U(n)\times.. \times U(n)$$ inducing reduced density state dependencies and associated tensor field splittings on multi-partite systems $\H\cong (\C^n)^{\otimes N>2}$.
As in the case of pure states, we believe that a connection to the IOVT-construction on Lie groups of general linear transformations (see \cite{Grabowski:2000zk}) should reduce the computational effort in concrete applications involving strata of mixed states with fixed rank. But also for more general submanifolds of quantum states, we may deal with the idea of computing quantum information distances on the level of a Hilbert space, rather than on the convex set of density states by taking into account quantum state purification procedures \cite{Man'ko:2000ti, Man'ko:2002ti}. In this way the advantage of dealing with probability amplitudes rather than with probability densities \cite{Facchi:2010} may be generalized from the regime of pure to mixed states. Besides the possible computational advantages related to density state purification, we shall also underline physical motivations for taking into account account pure rather mixed states as fundamental physical states \cite{Penrose:2004}. Hence, by the virtue of the latter point of view, we may in general put the geometry of the projective Hilbert space at the first place, even though we are dealing with the generalized regime of mixed states.

\section*{Acknowledgments}
This work was supported by the National Institute of Nuclear Physics (INFN).


\end{document}